\documentclass[9.5pt, nocopyrightspace]{sigplanconf}

\usepackage{eucal}
\usepackage{algorithm2e}
\usepackage{times,amsmath,amssymb,xspace}
\usepackage{txfonts}
\usepackage{graphicx}
\usepackage{multirow}
\usepackage{amsmath}
\usepackage{amssymb}
\usepackage{theorem}
\usepackage{epsfig}
\usepackage{fancyvrb}
\usepackage{url}
\usepackage{multirow}
\usepackage{wrapfig}
\usepackage{comment}
\usepackage{flushend}

\numberwithin{equation}{section}	

\theoremstyle{plain}			

\theoremstyle{definition}		

\newcommand{\Comment}[1]{}

\newcommand{\name}{\textsc{Cryptographic Path Hardening}\xspace}
\newcommand{\toolname}{\textsc{Cryptographic Path Hardener}\xspace}
\newcommand{\hard}{\mathcal{H}}

\begin{document}
\title{\name: \textsc{Hiding Vulnerabilities in Software through Cryptography}}

\authorinfo{
  Vijay Ganesh\and
  Michael Carbin\and
  Martin C. Rinard}
{Massachusetts Institute of Technology}
{\{vganesh, mcarbin, rinard\}@csail.mit.edu}

\pagestyle{plain}
\maketitle

\begin{abstract}

  We propose a novel approach to improving software security called
  \name, which is aimed at {\em hiding} security vulnerabilities in
  software from attackers through the use of provably secure and
  obfuscated cryptographic devices ~\cite{variaphd} to {\em harden}
  paths in programs.

  By ``harden'' we mean that certain error-checking {\it
    if}-conditionals in a given program $P$ are replaced by equivalent
  obfuscated {\it if}-conditionals in an obfuscated version of $P$. By
  ``hiding vulnerabilities'' we mean that adversaries cannot use
  semi-automatic program analysis techniques to reason about the {\it
    hardened program paths} and thus cannot discover as-yet-unknown
  errors along those paths, except perhaps through black-box
  dictionary attacks or random testing (which we can never prevent).
  Other than these unpreventable attack methods, we can make program
  analysis aimed at error-finding {\it provably hard} for a
  resource-bounded attacker, in the same sense that cryptographic
  schemes are hard to break. Unlike security-through-obscurity, in
  \name we use provably-secure crypto devices to hide errors and our
  mathematical arguments of security are the same as the standard ones
  used in cryptography.

  One application of {\name} is that software patches or filters often
  reveal enough information to an attacker that they can be used to
  construct error-revealing inputs to exploit an unpatched version of
  the program~\cite{brumley}. By {\it hardening} the patch we make it
  difficult for the attacker to analyze the patched program to
  construct error-revealing inputs, and thus prevent him from
  potentially constructing exploits.

\end{abstract}

\section{\name}

In \name we adopt the approach that one effective strategy for dealing
with security vulnerabilities is to make finding errors along hardened
paths in a program as computationally difficult as breaking some very
strong cryptographic assumption, such as the hardness of the discrete
log problem or factorization of large composite numbers whose factors
are large primes. We propose that we can achieve such guarantees by
developing a \toolname.

A \toolname takes as input a program P, and uses provably-secure
obfuscations, or generalizations thereof, to synthesize a path
hardened program $\hard(P)$ such that $\hard(P)$ has the following
properties:

\begin{enumerate}

\item {\bf Correctness:} $\hard(P)$ displays the same behavior as P on
  all inputs with very high probability

\item {\bf Polynomial Slowdown:} It is efficient to compile $P$ into
  $\hard(P)$, and also to run $\hard(P)$ on any input.

\item {\bf Security:} Parts of $\hard(P)$, such as certain kinds of
  conditionals, are obfuscated in a provably-secure manner.  As a
  consequence, program analysis of $\hard(P)$ aimed at constructing
  error-revealing inputs along hardened paths is {\it provably hard},
  in the same sense that cryptographic schemes are hard to break.

\end{enumerate}

A {\toolname} can synthesize $\hard(P)$ by identifying classes of
conditionals that can be re-implemented with off-the-shelf provably
obfuscated components.  For example, the if-conditional
\verb|if (x == a)| can be re-implemented with an obfuscated hash
function~\cite{1997_Canetti}, yielding an $\hard(P)$ that compares the
hash of \verb|x| with the pre-computed hash of \verb|a|.

\section{Case Study} 

A recent, high-profile vulnerability in PHP illustrates how a
{\toolname} could be used to quickly develop a hardened input filter
that can be distributed to protect an application and that, at the
same, does not reveal what type of input can exploit the application.

PHP 5.3.3 contains a vulnerability that can lead to denial-of-service
attacks on servers running web services implemented in
PHP~\cite{phpreport}.  The vulnerability is in PHP's routine for
converting the string representation of a decimal number into a
floating point value.  In particular, the routine computes the
floating point value via an iterative approximation algorithm; the
algorithm terminates when it reaches the floating point value that is
nearest to the decimal value of the string.

Due to the semantic differences between 80-bit extended precision
floating point registers and 64-bit IEEE doubles on 32-bit x86
architectures, this computation does not terminate when given a string
that represents the decimal number 2.2250738585072011e-308. Therefore,
if a malicious user passes such an input string to a vulnerable
server, then the PHP process will loop infinitely, consuming 100\% of
the available CPU resources.

PHP's developers eventually resolved this issue with a source patch,
but before the development of this resolution, some system
administrators publicly identified that they could quickly and
effectively block the attack with an input filter to their application
that rejected any web service request that contained the 128-bit
substring ``2250738585072011''~\cite{phpworkaround}. An astute
attacker could use this information to exploit the unpatched versions
of PHP, before the developers could release and fully deploy a source
patch.

\paragraph{A Hardened Filter. }

A {\toolname} can synthesize a hardened filter for this vulnerability
that satisfies our definitions by using one of a number of strong hash
functions (e.g., an obfuscated hash function ~\cite{1997_Canetti}, or
SHA-256).  Given such a hash function---which we denote by
\verb|hash|---we can construct a filter by pre-computing the hash of
the string ``2250738585072011''; let us denote this value by
\verb|s_hash|.  Given \verb|s_hash|, we can then implement the
substring check with a function that tests if each 16-byte substring
of the input matches \verb|s_hash|:

\begin{verbatim}
bool input_matches(string input) {
  for (int i = 0; i < length(input); ++i) {
    string str = input.substring(i, 16);
    if (hash(str) == s_hash)
      return true;
  }
  return false;
}
\end{verbatim}

This simple implementation satisfies each of three properties of a
path hardened program:

\paragraph {Correctness.} The implementation is correct with very high
probability.  There are two ways in which the implementation could be
incorrect: it could report that the input string contains the
malicious substring when it does not (a false positive), or it could
report that the input string does not contain the malicious substring
when it does (a false negative).  The second case does not occur when
the hash function is deterministic. The first case occurs with the
same probability that there is a collision in the selected hash
function. However, by selecting an appropriate hash function and hash
output length, we can make this probability very small or zero.

\paragraph {Polynomial Slowdown.} The implementation satisfies the
polynomial slowdown requirement of our definition. Both an unhardened
implementation and the hardened implementation of this filter run in
time linear in the length of the input string. However, the hardened
implementation will be some constant factor slower than an efficient
unhardened implementation because of the use of a hash function rather
than a simple bit-wise equivalence test.

\paragraph {Security.} The implementation is also secure in that an
attacker cannot determine the malicious substring without inverting
the hash function or guessing the 128-bit substring.

\subsection{Another Example} 

Another simple generalization of the above application of {\name} is
to harden patches or filters where the filter checks inputs against a
{\it small} set or range of values ($a <= x <= b$).  Here {\it small}
refers to the fact that the number of error-triggering values checked
by the filter is much smaller than the total number of values that the
input variables can take. It is also assumed, in this context, that
these error-triggering values are difficult to guess. For example,
below is a patch that checks if the input variable $x$ can take any
value from a small set of values $v_i$. If the answer is YES, then
reject the input else accept. Let $C$ denote the disjunctive
conditional $\bigvee x == v_i$:

\begin{verbatim} 
 if (C) { exit (1);}
\end{verbatim}

Such a filter is a prime candidate for {\name}. Specifically,
distributing a patch that reveals the exact semantics of the
conditional would give attackers exactly the condition they need to
exploit the vulnerability. Moreover, a {\toolname} can easily
construct a hardened implementation of the filter by disjunctively
comparing the hash of the input variable $x$ with pre-computed hashes
of $v_i$. Let $hard(C)$ denote $\bigvee hash(x) == hash\_{v_i}$, where
$hash\_v_i$ are the pre-computed hash values of the $v_i$'s. Then the
hardened filter is:

\begin{verbatim}
 if (hard(C)) { exit (1);}
\end{verbatim}

\section{Discussion} 

\name has a number of points of discussion concerning its applicability
and practicality.

\paragraph{Hardness of Inversion.} The notion of \name is well-defined
for all conditionals, but it is only meaningful for conditionals that
are difficult to dictionary attack in a black-box manner.
Specifically, if the conditional $\phi$ has the form that it is easy
for an adversary to find a satisfying assignment given only a black
box that implements $\phi$, then there is no hope of hardening $\phi$.
For example, if $\phi(x)$ is the conditional that implements an
inequality check of the form $0 < x < c$ or $MAX > x > c$ for some
constant $c$, where the range of values of $x$ checked by the
inequality is {\it large}, then it is easy to find a satisfying
assignment (and indeed, $c$) by binary search.  Similarly, if $\phi$
is a conditional that checks whether a small-sized substring (say, one
character) is present in its input string, then it would also be
simple to find a satisfying assignment just by searching through all
possible characters (which is a small enough set that it is quick to
do).

\paragraph{Correctness.}  According to our definition, a target
conditional and its hardened counterpart may semantically differ. In
particular, our definition accepts hardened implementations that may,
with some probability, evaluate to true in instances where the
original conditional evaluates to false (i.e.,
false-positives). Therefore, a user of a {\toolname} must be able
reason about whether it's acceptable for the resulting hardened
program to be overly conservative in, for example, rejecting inputs.

For simple conditionals that can be directly implemented with hash
functions (such as testing a variable against a constant), the
probability of error is equivalent to the probability of collisions
for the chosen hash function.

\paragraph{Performance.} A cryptographic implementation of a
conditional can be slower than its standard implementation.  In
general, standard performance analysis considerations must be taken in
account when applying \name. For example, hardening conditionals on
hot loops may incur more whole-program performance slowdown than
hardening conditionals in infrequently executed code.

\section{Conclusion}
In this paper, we demonstrated how {\name} can be used to enable
developers to distribute hardened patches that hide the exact
conditions of an exploit. Furthermore, we propose that hiding
vulnerabilities through {\name} presents a new approach to software
security that is paradigmatically different from traditional software
engineering approaches such as {\it formal methods} that focus on
proving the absence of errors in programs, or {\it testing techniques}
that focus on establishing the presence of errors. 

A possible generalization of the ideas presented here could be to use
{\em hard-to-invert} functions to hide all {\em non-observable}
constants in a program P, thus slowing down any adversary who may want
to use analysis to find as-yet-unknown errors in P.



\bibliographystyle{abbrv}
\bibliography{proghard-short}
\end{document}